# An Instance-based Plus Ensemble Learning Method for Classification of Scientific Papers

Fang Zhang[1] • Shengli Wu[2]
1 School of Education, Hubei University of Arts and Science, Xiangyang, China
2 School of Computing, Ulster University, Belfast, UK

**ABSTRACT.** The exponential growth of scientific publications in recent years has posed a significant challenge in effective and efficient categorization. This paper introduces a novel approach that combines instance-based learning and ensemble learning techniques for classifying scientific papers into relevant research fields. Working with a classification system with a group of research fields, first a number of typical seed papers are allocated to each of the fields manually. Then for each paper that needs to be classified, we compare it with all the seed papers in every field. Contents and citations are considered separately. An ensemble-based method is then employed to make the final decision. Experimenting with the datasets from DBLP, our experimental results demonstrate that the proposed classification method is effective and efficient in categorizing papers into various research areas. We also find that both content and citation features are useful for the classification of scientific papers.

## 1  Introduction

Classification of scientific papers is an important task that can be useful in various ways. It can be used for effective and efficient search of scientific papers. It can also be used for other tasks, for example to analyze the tendency of a research topic or a research field. It can be beneficial for researchers to identify relevant or high-impact papers and research topics. It can be used for research institutions and government agencies to understand the trends in each research field to help set research and funding policies, among other applications.

Up to now, there are two types of approaches for this research problem. One is the manual plus machine approach (Kandimalla et al., 2020; Mendoza et al., 2022; Zhang et al., 2022), and the other is a completely automatic machine-based approach (Shen et al., 2018; Toney-Wails & Dunham, 2022; Zhang et al., 2022, Liu et al., 2022). For the manual plus machine approach, we first define a classification system, which is a hierarchical structure including all the categories and sub-categories into which all scientific papers should fit. A classification system also provides specifications of those categories and sub-categories involved. On the other hand, for a given set of scientific papers, a completely automatic machine-based approach tries to work out a hierarchical structure and all corresponding papers belonging to each category by an algorithm. Understandably, the second approach is more ambitious and challenging than the first one.

In this paper, we focus on the first approach. With a given classification system and a number of manually identified papers for each category, we try to classify all incoming papers in an effective and efficient manner. The main contributions of this work are as follows:

1. We propose a novel approach that employs instance-based learning and ensemble learning. First by using instance-based learning with content related features, papers are classified into specific research areas. Papers are also classified by using citations in both directions. Afterwards, an ensemble-based method is employed to make the final decision.
2. Experiments with the DBLP datasets demonstrate that the proposed method can achieve very good classification results.

## 2 Related Work

In this section we review some recent work on the classification of scientific papers. We also briefly review some major methods in ensemble learning.

### 2.1 Classification of scientific papers

The classification of scientific papers has become an important issue when organizing and managing an increasing number of publications through computerized solutions. In previous research, typically, meta-data such as title, abstract, keywords, and citations of papers were used for this task, while full text was not considered due to its unavailability in most situations.

Various machine learning methods, such as K-Nearest Neighbors (Waltman and Van Eck, 2012; Lukasik et al., 2013), K-means (Kim and Gil, 2019), and Naïve Bayes (Eykens et al., 2021), have been applied. Recently, deep neural network models, such as Convolutional Neural Networks (Rivest et al., 2021; Daradkeh et al., 2022), Recurrent Neural Networks (Semberecki and Maciejewski, 2017; Hoppe et al., 2021), and pre-trained language models (Kandimalla et al., 2020; Hande et al., 2021), have also been utilized.

One key issue is the classification system to be used. There are many different classification systems. Both Thomson Reuters' Web of Science database (WoS) and Elsevier's Scopus database have their own general classification systems, covering many subjects/research areas. Some systems focus on one particular subject, such as the Medical Subject Headings (MeSH), the Physics and Astronomy Classification Scheme (PACS), the Chemical Abstracts Sections, the Journal of Economic Literature (JEL), and the ACM Computing Classification System.

Based on the WoS classification system, Kandimalla et al. (2020) applied a Deep Attentive Neural Network (DANN) to a collection of papers from the WoS database for the classification task. It was assumed that each paper belonged to only one category, and only abstracts were used.

Zhang et al. (2022) compared three classification systems: Thomson Reuters' Web of Science, Fields of Research provided by Dimensions, and the Subjects Classification provided by Springer Nature. Among these, the second one was generated by machine learning methods automatically, while the other two were generated manually by human experts. It is found there is significant inconsistency between machine and human-generated systems.

Rather than using an existing classification system, some researchers build their own classification system using the collection to be classified or other resources such as Wikipedia.

Shen et al. (2018) organized scientific publications into a hierarchical concept structure of up to six levels. The first two levels (similar to areas and sub-areas) were manually selected, while the others were automatically generated. Wikipedia pages were used to represent the concepts. Each publication or concept was represented as an embedding vector, thus the similarity between a publication and a concept could be calculated by the cosine similarity of their vector representations. It is a core component for the construction of the Microsoft Academic Graph.

In the same vein as Shen et al. (2018), Toney-Wails & Dunham (2022) also used Wikipedia pages to represent concepts and build the classification system. Both publications and concepts were represented as embedding vectors. Their database contains more than 184 million documents in English and more than 44 million documents in Chinese.

Mendoza et al. (2022) presented a benchmark corpus and a classification system as well, which could be used for the academic paper classification task. The classification system used is the 36 subjects defined in the UK Research Excellent Framework. According to Cressey & Gibney (2014), this practice is the largest overall assessment of university research outputs ever undertaken globally. The 191,000 submissions to REF 2014 comprise a very good data set because every paper was manually categorized by experts when submitted.

Liu et al. (2022) described the NLPCC 2022 Task 5 Track 1, a multi-label classification task for scientific literature, where one paper may belong to multiple categories simultaneously. The data set, crawled from the American Chemistry Society's publication website, comprises 95,000 papers' meta-data including titles and abstracts. A hierarchical classification system, with a maximum of three levels, was also defined.

As we can see, the classification problem of academic papers is quite complicated. Many classification systems and classification methods are available. However, classification systems and classification methods are related to each other. The major goal of this work is to perform citation count prediction of published papers, in which classification of papers is a basic requirement. For example, considering the DBLP dataset which includes over four million papers, special consideration is required to perform the classification task effectively and efficiently. We used the classification system from CSRankings, which included a set of four categories (research areas) and 26 sub-categories in total. A group of top venues were identified for each sub-category. However, many more venues in DBLP are not assigned to any category. We used all those recommended venue papers in the CSRankings system as representative papers of a given research area. An instance-based learning approach was used to measure the semantic similarity of the target paper and all the papers in a particular area. A decision could be made based on the similarity scores that the target paper obtained for all research areas. Besides, citation data between the target paper and all the papers in those recommended venues is also considered. Quite different from those proposed classification methods before, this instance-based learning approach suits our purpose well. See Section 3 for more details.

**2.1 Ensemble learning**

Ensemble learning has been investigated by many researchers over the last thirty years for classification tasks (Zhou, 2012). Mainly ensemble learning approaches can be categorized into three typical types: stacking (Wolpert, 1992), bagging (Breiman, 1996), and boosting (Opelt & Pinz, 2006). Staking is focused on combining different types of base learners to achieve better results. A few different types of combination schemes have been proposed: Stacking (Ting & Witten, 1999), StackingC (Seewald, 2002), and Euclidean Distance-based method (Wu et al., 2023). Bagging is focused on generating a group of diversified base learners through instance or feature-level manipulation of a dataset (Breiman, 2001; Xu &Yu, 2023). Boosting is also focused on generating a group of useful base learners, however, in a different way from Bagging. It generates base classifiers one after another and more and more focused on those instances that have been classified incorrectly. AdaBoost (Freund & Schapire, 1995) and XGBoost (Chen & Guestrin, 2016) are two representatives. In this piece of work, we used the Euclidean Distance-based method. It fits our purpose well.

Similarly, ensemble learning is referred to as data fusion in information retrieval [Huang & Xu, 2022]. The primary principle remains the same for both of them. Although it seems that classification and regression are the main problems in machine learning, while ranking is the main problem in information retrieval, they can be transformed from one type to another.

# 3 Methodology

**3.1 Computing classification system**

To carry out the classification task of academic papers, a suitable classification system is required. There are many classification systems available for natural science, social science, humanities, or specific branches of science or technology. Since one of the datasets used in this study is DBLP, which includes over four million papers on computer science so far, we will focus our discussion on classification systems and methods for computer science.

In computer science, there are quite a few classification systems available. For example, both the Association for Computing Machinery (ACM) and the China Computer Federation (CCF) define their own classification systems. However, neither are suitable for our purpose. The ACM's classification system is quite complicated, but it does not provide any representative venues for any of the research areas. The CCF defines 10 categories and recommends dozens of venues in each category. However, some journals and conferences publish papers in more than one category, but they are only recommended in one category. For instance, both the journals IEEE Transactions on Knowledge and Data Engineering and Data and Knowledge Engineering publish papers on Information Systems and Artificial Intelligence, but they are only recommended in the Database/Data Mining/Content Retrieval category. In this research, we used the classification system from CSRankings. This system divides computer science into four areas: AI, System, Theory, and Interdisciplinary Areas. Then, each area is further divided into several sub-areas, totaling 26 sub-areas. See Appendix for the list of those sub-areas. We flatten these 26 sub-areas for classification, while ignoring the four general areas at level one. One benefit of using this system is that it lists several key venues for every sub-area. For example, three venues are identified for Computer Vision: CVPR (IEEE Conference on Computer Vision and Pattern Recognition), ECCV (European Conference on Computer Vision), and ICCV (IEEE International Conference on Computer Vision). This is very useful for the paper classification task, as we will discuss now.

### 3.2 Paper classification

For this research, we need a classification algorithm that can perform the classification task for all the papers in the DBLP dataset effectively and efficiently.

Although many classification methods have been proposed, we could not find a method that suits our case well. Therefore, we developed our own approach. Using the classification system of CSRankings, we assume that all the papers published in those identified venues belong to that given research area, referred to as seed papers. For all the non-seed papers, we need to decide the areas to which they belong. This is done by considering three aspects together: content, references, and citations. Let us look at the first aspect first.

The collection of all the seed papers, denoted as $C$, was indexed using the Search engine Lucene [1] with the BM25 model. Both titles and abstracts were used in the indexing process. Each research area $a_k$ is presented by all its seed papers $C(a_k)$. For a given non-seed paper $p$, we use its title and abstract as a query to search for similar papers in $C$. Then each seed paper $s$ will obtain a score (similarity between $s$ and $p$)

$$sim(p,s) = \sum_{t_j \in P} idf(q_j) \times \frac{f(t_j,s) \times (k_1 + 1)}{f(t_j,s) + c_1 \times \left(1 - b + b \times \frac{|S|}{avg\_d\_l}\right)} \quad (1)$$

in which $b$ and $k_1$ are two parameters (set to 0.75 and 1.2, respectively, as default setting values of Lucene in the experiments), $S$ is the set of all the terms in $s$, $P$ is the set of all the terms in $p$, $avg\_d\_l$ is the average length of all the documents in $C$, $f(t_j,s)$ is the term frequency of $t_j$ in $s$, $idf(t_j)$ is the inverse document frequency of $t_j$ in

---
[1] https://lucene.apache.org

collection $C$ with all the seed papers. $idf(t_j)$ is defined as

$$idf(t_j) = \log\left(1 + \frac{|C| - f(t_j) + 0.5}{f(t_j) + 0.5}\right) \quad (2)$$

in which $|C|$ is the number of papers in $C$, and $f(t_j)$ is the number of papers in $C$ satisfying the condition that $t_j$ appears in them. For a paper $p$ and a research area $a_k$, we can calculate the average similarity score between $p$ and all the seed papers in $C(a_k)$ as

$$sim(p, a_k) = \frac{1}{|C(a_k)|} \sum_{s \in C(a_k)} sim(p, s) \quad (3)$$

where $C(a_k)$ is the collection of seed papers in $a_k$.

We also consider citations between $p$ and any of the papers in $C$. Citations in two different directions are considered separately: $citingNum(p, a_k)$ denotes the number of papers in $C(a_k)$ that $p$ cites, and $citedNum(p, a_k)$ denotes the number of papers in $C(a_k)$ that cites $p$. Now we want to combine the three features. Normalization is required. For example, $sim(p, a_k)$ can be normalized by

$$sim'(p, a_k) = \frac{sim(p, a_k)}{\sum_{a_l \in RArea} sim(p, a_l)} \quad (4)$$

in which $RArea$ is the set of 26 research areas. $citingNum'(p, a_k)$ and $citedNum'(p, a_k)$ can be normalized similarly. Then we let

$$score(p, a_k) = \beta_1 \times sim'(p, a_k) + \beta_2 \times citingNum'(p, a_k) + \beta_3 \times citedNum'(p, a_k) \quad (5)$$

where $C(a_k)$ is the collection of seed papers in $a_k$.

for any $a_k \in RArea$, in which $\beta_1$, $\beta_2$, and $\beta_3$ are three parameters. When applying Equation 5 to $p$ and all 26 research areas, we may obtain corresponding scores for each area. $p$ can be put to research area $a_k$ if $score(p, a_k)$ is the biggest among all 26 scores for all research areas. The values of $\beta_1$, $\beta_2$, and $\beta_3$ are decided by Euclidean Distance with multiple linear regression with a training data set (Wu et al., 2023). Compared with other similar methods such as Stacking with MLS and StackingC, this method can achieve comparable performance but much more efficient than the others. It should be very suitable for large-scale datasets.

In this study, we assume that each paper just belongs to one of the research areas. If required, this method can be modified to support multi-label classification, then a paper may belong to more than one research area at the same time. We may set a reasonable threshold $\tau$, and for any testing paper $p$ and research area $a$, if $score(p, a_k) > \tau$，then paper $p$ belongs to research area $a$. However, this is beyond the scope of this research, and we leave it for further study.

In summary, the proposed classification algorithm IBL (Instance-Based Learning) is sketched as follows:

---

**ALGORITHM 1**: IBL

---

**Input**: a set of research areas $a_k$ $(1 \leq k \leq n)$, each includes a group of papers that belong to that area, a paper $p$ to be classified, citation data of $p$, and parameter $\beta_1, \beta_2, \beta_3$, for the regression model

**Output**: the research area that $p$ belongs to

1  **For** every research areas $a_k$ $(1 \leq k \leq n)$
2    calculate the average similarity $sim'(p, a_k)$ between p and every paper in $a_k$ by Equation 4
3    calculate normalized citation count that $p$ cites $citingNum'(p, a_k)$
4    calculate normalized citation count that $p$ is been cited $citedNum'(p, a_k)$
5    calculate $score(p, a_k)$ by Equation 5
6  **End For**
7  compare all the scores that $p$ obtains, return the research area $a_i$ that $p$ obtains the maximal score

---

## 4 Experimental Settings and Results

### 4.1 Dataset and measures

We downloaded a DBLP dataset (Tang et al., 2008)[2]. It contains 4,107,340 papers in computer science and 36,624,464 citations from 1961 to 2019. For every paper, the dataset provides its metadata such as title, abstract, references, authors and their affiliations, publication year, the venue in which the paper was published, and citations since publication. Some subsets of it were used in this study.

We used two subsets of the dataset. The first one ($S_1$) is all the papers published in those 72 recommended venues in CSRankings between 1965 and 2019. There are 191,727 papers. $S_1$ is used as seed papers for all 26 research areas. The second subset ($S_2$) includes 1300 papers, 50 for each research area. Those papers were randomly selected from a group of 54 conferences and journals and judged manually. $C_2$ is used for the testing of the proposed classification method.

Both accuracy and F-measure are used for the evaluation of the classification results.

### 4.2 Classification results

In the CSRankings classification system, there are a total of 26 special research areas. A few top venues are recommended for each of them. We assume that all the papers published in those recommended conferences belong to the corresponding research area solely. For example, three conferences CVPR, ECCV, and ICCV are recommended for Computer Vision. We assume that all the papers published in these three conferences belong to the Computer Vision research area but no others.

To evaluate the proposed method, we used a set of 1300 non-seed papers ($S_2$). It included 50 papers for each research area. All of them were labelled manually. In Equation 5, three parameters need to be trained. Therefore, we divided those 1300 papers into two equal partitions of 650, and each included the same number of papers in every research area. then a two-fold cross-validation approach was performed. Table 1 shows the average performance.

---

[2] https://www.aminer.cn/

We can see that the proposed method with all three features, content similarity ($sim'$), citation to other papers ($citingNum'$), and citation by others ($citedNum'$), are useful for the classification task. Roughly citation in both directions ($citingNum'$ + $citedNum'$) and content similarity ($sim'$) have the same ability. Considering three features together, we can obtain an accuracy, or an F-measure, of approaching 0.8. We are satisfied with this solution. On the one hand, its classification performance is good compared with other methods in the same category, e.g., (Kandimalla et al., 2020, Ambalavanan & Devarakonda, 2020). In Kandimalla et al. (2020), F-scores across 81 subject categories are between 0.5-0.8 (See Figure 5 in that paper). In Ambalavanan & Devarakonda (2020), the four models ITL, Cascade Learner, Ensemble-Boolean, and Ensemble-FFN obtain an F-score of 0.553, 0.753, 0.628, and 0.477, respectively, on the Marshall dataset they experimented with (see Table 4 in their paper). Although those results may not be comparable since the datasets used are different, it is an indicator that our method is very good. Besides, our method can be implemented very efficiently. When the seed papers are indexed, we can deal with a large collection of papers very quickly with very little resource. The method is very scalable.

**Table 1** Performance of the classification method and its features

| Method | Accuracy | F-measure |
| --- | --- | --- |
| $citingNum'$ | 0.377 | 0.443 |
| $citedNum'$ | 0.490 | 0.542 |
| $sim'$ | 0.645 | 0.636 |
| $citingNum' + citedNum'$ | 0.622 | 0.652 |
| $citingNum' + sim'$ | 0.716 | 0.711 |
| $citedNum' + sim'$ | 0.752 | 0.746 |
| All three features | **0.795** | **0.790** |

Finally, we set up a linear regression model to investigate the impact of the three factors to classification performance by using them as independent variables while classification performance (either Accuracy or F-measure) as the dependent variable. It is found that all three factors impact classification performance significantly at the level of 95%.

## 5  Conclusions

In this paper, we have presented a novel instance-based and ensemble learning method for the classification of scientific papers. Both content and citations in both directions are considered in the classification model at the same time. Experimented with the datasets from DBLP, we find that the proposed method is effective and efficient. All three factors impact classification performance significantly.

This piece of work can be further investigated in a few different directions. Firstly, some papers may be of the nature of cross-disciplinary. It would be better to allocate such papers to more than one category. Secondly, some other information such as authors and venues are also useful information, which can be employed to improve classification performance. Some graph-based machine learning methods can be good options for dealing with such situations.

# Appendix

**Table 1.** All Areas and their sub-areas in the CSRankings classification system for papers in computer science

| 1. AI | 3. Theory |
|---|---|
| 1.1 Artificial intelligence | 3.1 Algorithms & complexity |
| 1.2 Computer vision | 3.2 Cryptography |
| 1.3 Machine learning | 3.3 Logic & verification |
| 1.4 Natural language processing | |
| 1.5 The Web & information retrieval | |
| | |
| **2. Systems** | **4. Interdisciplinary Areas** |
| 2.1 Computer architecture | 4.1 Comp. bio & bioinformatics |
| 2.2 Computer networks | 4.2 Computer graphics |
| 2.3 Computer security | 4.3 Computer science education |
| 2.4 Databases | 4.4 Economics & computation |
| 2.5 Design automation | 4.5 Human-computer interaction |
| 2.6 Embedded & real-time systems | 4.6 Robotics |
| 2.7 High-performance computing | 4.7 Visualization |
| 2.8 Mobile computing | |
| 2.9 Measurement & perf. analysis | |
| 2.10 Operating systems | |
| 2.11 Programming languages | |